\documentclass[onecolumn,showpacs,prd]{revtex4}
\usepackage{bm}
\usepackage{hyperref}
\usepackage{epsfig}

\newcommand{\NPA}[3]{Nucl.\ Phys.\ {\bf A#1}, #2 (#3)}

\newcommand{\PLB}[3]{Phys.\ Lett.\ B\ {\bf #1}, #2 (#3)}

\newcommand{\PRL}[3]{Phys.\ Rev.\ Lett.\ {\bf #1}, #2 (#3)}

\newcommand{\PRC}[3]{Phys.\ Rev.\ C\ {\bf #1}, #2 (#3)}
\newcommand{\PRD}[3]{Phys.\ Rev.\ D\ {\bf #1}, #2 (#3)}
\newcommand{\JPG}[3]{J.\ Phys.\ G\ {\bf #1}, #2 (#3)}

\begin{document}

\title{Landau levels of cold dense quark matter in a strong magnetic field}

\author{
Xin-Jian Wen\footnote{wenxj@sxu.edu.cn}, Jun-Jun
Liang\footnote{liangjj@sxu.edu.cn} }

\affiliation{ Institute of Theoretical Physics, Shanxi University,
Taiyuan 030006, China
              }
\date{\today}

\begin{abstract}
The occupied Landau levels of strange quark matter are investigated
in the framework of the SU(3) NJL model with a conventional coupling
and a magnetic-field dependent coupling respectively. At lower
density, the Landau levels are mainly dominated by $u$ and $d$
quarks. Threshold values of the chemical potential for the $s$ quark
onset are shown in the $\mu$-$B$ plane. The magnetic-field-dependent
running coupling can broaden the region of three-flavor matter by
decreasing the dynamical masses of $s$ quarks. Before the onset of
$s$ quarks, the Landau level number of light quarks is directly
dependent on the magnetic field strength $B$ by a simple inverse
proportional relation $k_{i,\mathrm{max}}\approx B_i^0/B$ with
$B_d^0=5\times 10^{19}$ G, which is  approximately 2 times $B_u^0$
of $u$ quarks at a common chemical potential. When the magnetic
field increases up to $B^0_d$, almost all three flavors are lying in
the lowest Landau level.
\end{abstract}
\pacs{12.39.-x, 12.40.jn, 12.38.Mh} \maketitle
\section{INTRODUCTION}
The study of Quantum Chromodynamics (QCD) matter subject to a strong
magnetic field has been a hot topic  of intense investigation
\cite{Miransky15}. The strange quark matter (SQM) is regarded as a
ground state composed of deconfined $u$, $d$, and $s$ quarks
\cite{Witten}. The new state is expected to be searched in extreme
conditions of high density and/or high temperature. In addition to
these environments, the SQM is argued to be subject to strong
magnetic fields. The extreme strong magnetic field theoretically
seems beyond the scope of the conventional condensed matter, and its
origin remains not very clear until now. However, it has been
recently proposed to be produced in noncentral collision experiments
in the Relativistic Heavy Ion Collider and the Large Hadron Collider
on the one hand \cite{magnetic,skok09}, or to be naturally existing
in the core of pulsars on the other hand. The large magnetic fields
in nature are normally associated with astrophysical objects, where
the density is much higher than the nuclear saturation. The typical
strength could be the order of $10^{12}$ G on the surface of pulsars
\cite{Dong91}. Some magnetars can have even larger magnetic fields,
reaching the surface value as large as $10^{14}- 10^{15}$ G
\cite{thom92}. By comparing the magnetic and gravitational energies,
the physical upper limit to the total neutron star is of the order
$10^{18}$ G. For self-bound quark stars, the limit could go higher
\cite{Chanm01}. Maximum strengths of $10^{18}- 10^{20}$ G in the
interior of stars are proposed by an application of the viral
theorem \cite{magnetic,Dong91}. In the LHC/CERN energy, it is
possible to produce a field as large as $5\times 10^{19}$ G
\cite{skok09}.

The special properties of QCD matter are widely affected by strong
magnetic fields in many branches, such as the (inverse) magnetic
catalysis \cite{klimenko,catapaper,FerrerPRL11,Ferreira14,Huang15},
the anisotropies \cite{anisotr05,Meneze15}, the magnetic
oscillations \cite{Ebert99}, the magnetization \cite{Meneze15}, the
phase diagram with a critical end point \cite{Avan12,costa15} etc.
The magnetic field larger than $10^{19}$ G can obviously change the
spherical symmetry \cite{Felipe09}. All of these are essentially
resulted due to the Landau levels arrangement of charged particles
in magnetic fields. In principle, not only quark masses will change
in the medium, but also the coupling constant will run in the
medium, such as the magnetic-field-dependent coupling and
magnetic-temperature-dependent coupling \cite{Farias}. It is well
known that the dressed masses of three flavors are very different,
which leads to a flavor-dependent fraction in quark matter and
strangelets \cite{wen2005}. Similarly, the threshold condition and
the quantum numbers of the Landau level of quarks are also flavor
dependent. Generally, $u$ and $d$ quarks dominate the bulk matter at
low densities. As the density increases, strange flavor begins to
take part in its Landau levels. Therefore, the phase was argued to
be divided into three regions such as the chirally broken phase
(B-phase), the massive phase (C-phase), and the chirally restored
phase (A-phase) in previous work \cite{allen13,Grunfeld}, where the
detailed locations of the two-flavor Landau Levels were shown in the
$\mu$-$B$ phase panel. However, at a proper density, the $s$ quark
cannot participate in the previous discussion because its dressed
mass is larger than the chemical potential, thus the $s$ quark could
not occupy its lowest Landau level (LLL). In this work, we will show
the critical density for the appearance of $s$ quarks by considering
the two kinds of coupling interactions, the conventional coupling
constant and the magnetic-field-dependent running coupling
respectively. The main aim of this work is to perform a detailed
analysis of the Landau levels with and without the $s$ quark
depending on the magnetic field strength .

This work is organized as follows. In Sec. \ref{sec2}, a brief
review of the Nambu$-$Jona-Lasinio (NJL) model description of cold
SQM in a strong magnetic field is provided. The
magnetic-field-dependent running scalar coupling in the SU(3)
version is introduced as well as the model parameters in the
computation. In Sec. \ref{sec:result}, the numerical results and
discussion are given at common chemical potential and under the
$\beta$ equilibrium respectively. A detailed analysis of the
occupied Landau Levels with respect to the magnetic field is given.
The last section is a short summary.

\section{Thermodynamics of Magnetized SQM In the SU(3) NJL model}
\label{sec2}

The SU(3) NJL Lagrangian density includes both a scalar-pseudoscalar
interaction and the t'Hooft six-fermion interaction \cite{hats94}
and can be written as \cite{Grunfeld},
\begin{equation}{\cal{L}}_{NJL}=\bar{\psi}( i/\kern-0.5em D-m )\psi
+G\sum_{a=0}^8[(\bar{\psi}\lambda_a \psi)^2
+(\bar{\psi}\gamma_5\lambda_a
\psi)^2]-K\{\det[\bar{\psi}(1+\gamma_5)\psi]+\det[\bar{\psi}(1-\gamma_5)\psi]\}.
\end{equation}
The field $\psi=(u, d, s)^T$ represents a quark field with three
flavors. Correspondingly, $m=\mathrm{diag}(m_u, m_d, m_s)$ is the
current mass matrix with $m_u=m_d\neq m_s$. $\lambda_0=\sqrt{2/3}I$
where $I$ is the unit matrix in the three-flavor space. $\lambda_a$
with $0<a\leq 8$ denotes the Gell-Mann matrix. The gap equations for
three-flavor are coupled and should be solved consistently,
\begin{eqnarray}M_i-m_i+4G\phi_i-2K\phi_j\phi_k=0, \label{eq:gapSU3}
\end{eqnarray}
where $(i$, $j$, $k)$ is the permutation of $(u$, $d$, $s)$. The
contribution from the quark flavor $i$ is
\begin{equation}\phi_i=\phi_i^\mathrm{vac}+\phi_i^\mathrm{mag}+\phi_i^\mathrm{med}.\label{eq:condensate}
\end{equation}

The terms $\phi_i^\mathrm{vac}$, $\phi_i^\mathrm{mag}$, and
$\phi_i^\mathrm{med}$ representing the vacuum, magnetic field, and
medium contribution to the quark condensation are respectively
\cite{menezes09}
\begin{eqnarray}\phi_i^\mathrm{vac}&=&-\frac{M_i N_c}{2\pi^2} [\Lambda
\sqrt{\Lambda^2+M_i^2}-M_i^2\ln(\frac{\Lambda+\sqrt{\Lambda^2+M_i^2}}{M_i})],
\\
\phi_i^\mathrm{mag}&=& -\frac{M_i |q_i| B N_c}{2\pi^2} \left\{
\ln[\Gamma(x_i)]-\frac{1}{2}\ln(2\pi)+x_i-\frac{1}{2}(2x_i-1)\ln(x_i)
\right\},
\\
\phi_i^\mathrm{med}&=& \sum_{k_i=0}^{k_{i,\mathrm{max}}} a_{k_i}
\frac{M_i|q_i|BN_c}{2\pi^2} \ln\left[
\frac{\mu_i+\sqrt{\mu_i^2-s_i^2}}{s_i} \right].
\end{eqnarray}
The effective quantity $s_i=\sqrt{M_i^2+2k_i|q_i|B}$ sensitively
depends on the magnetic field. The dimensionless quantity is
$x_i=M_i^2/(2|q_i|B)$. The degeneracy label of the Landau energy
level is $a_{k_i}=2-\delta_{k0}$. The quark condensation is greatly
strengthened by the factor $|q_iB|$ together with the dimension
reduction $D-2$ \cite{Miransky,Kojo14}. The Landau quantum number
$k_i$ and its maximum $k_{i,\mathrm{max}}$ are defined as
\begin{equation}k_i\leq k_{i,\mathrm{max}}=
\mathrm{Int}[\frac{\mu_i^2-M_i^2}{2|q_i|B}],
\end{equation} where ``Int'' means the number before the decimal point.

The total thermodynamic potential density in the mean field
approximation reads
\begin{equation}\label{omega}\Omega=\sum_{i=u,d,s}(\Omega_i^\mathrm{vac}+\Omega_i^\mathrm{mag}+\Omega_i^\mathrm{med}+2G\phi^2_i) -4K \phi_u\phi_d\phi_s.
\end{equation}
where the first term in the summation is the vacuum contribution to
the thermodynamic potential, i.e.,
\begin{eqnarray}\Omega_i^\mathrm{vac}= \frac{N_c}{8 \pi^2}
\left[M_i^4\ln(\frac{\Lambda+\epsilon_\Lambda}{M_i})-\epsilon_\Lambda
\Lambda (\Lambda^2+\epsilon_\Lambda^2) \right],
\end{eqnarray}where the quantity $\epsilon_\Lambda$ is defined as
$\epsilon_\Lambda=\sqrt{\Lambda^2+M_i^2}$. The ultraviolet
divergence in the vacuum part of the thermodynamic potential
$\Omega$ is removed by the momentum cutoff. In the literature, a
form factor is introduced in the diverging zero energy as a smooth
regularization procedure \cite{Gatto2010}. The magnetic field and
medium contributions are respectively
\begin{eqnarray}
\Omega_i^\mathrm{mag}&=& -\frac{N_c (|q_i|B)^2}{2 \pi^2} \left[ \zeta'(-1,x_i)-\frac{1}{2}(x_i^2-x_i)\ln(x_i)+\frac{x_i^2}{4} \right], \\
\Omega_i^\mathrm{med}&=& -\frac{|q_i|B N_c}{4\pi^2}
\sum_{k=0}^{k_\mathrm{max}} a_{k_i} \left\{ \mu_i
\sqrt{\mu_i^2-(M_i^2+2k_i |q_i|B)}-(M_i^2+2k_i
|q_i|B)\ln[\frac{\mu_i+\sqrt{\mu_i^2-(M_i^2+2k_i
|q_i|B)}}{\sqrt{M_i^2+2k_i |q_i|B}}] \right\}\label{eq:ome_med},
\end{eqnarray} where
$\zeta(a,x)=\sum_{n=0}^\infty \frac{1}{(a+n)^x}$ is the Hurwitz zeta
function. From the thermodynamic potential (\ref{omega}), one can
easily obtain the quark density as
\begin{eqnarray}
n_i(\mu,B)=\sum_{k=0}^{k_{i,\mathrm{max}}} a_{k_i} \frac{|q_i|
BN_c}{2\pi^2} \sqrt{\mu_i^2-(M_i^2+2k_i |q_i|B)}.
\end{eqnarray} The corresponding pressure from the flavor $i$ contribution is
\begin{eqnarray}P_i(\mu_i,B)=-\Omega_i=-(\Omega_i^\mathrm{vac}+\Omega_i^\mathrm{mag}+\Omega_i^\mathrm{med}).
\end{eqnarray}
Under strong magnetic fields, the system total pressure should be a
sum of the matter pressure and the field pressure contribution
\cite{menezes09,maxwell}. So we have
\begin{eqnarray}P_i(\mu_i,B)=-\Omega_i+\frac{B^2}{2},
\end{eqnarray}where the magnetic field term $B^2/2$ is due to the
electromagnetic Maxwell contribution. It is well known to us that
the energy density and pressure should vanish in vacuum. So the
pressure and the thermodynamic potential should be normalized by
requiring the zero pressure at the zero density as \cite{menezes09}
\begin{eqnarray}\label{eq:normali}P_i^\mathrm{eff}(\mu_i,B) = P_i(\mu_i,B)-P_i(0,B).
\end{eqnarray}In the normalization result, the field term is automatically absent.
According to the fundamental thermodynamic relation, the free energy
density at zero temperature is
\begin{eqnarray} \varepsilon_i =
-P_i^\mathrm{eff}+\mu_i n_i.
\end{eqnarray}

The system pressure and energy density are written as
\begin{eqnarray}P=\sum_i P_i^\mathrm{eff}, \ \ \ \ \
\varepsilon=\sum_i \varepsilon_i,
\end{eqnarray}where the summation goes over $u$, $d$ quarks, and electrons.

In principle, the interaction coupling constant between quarks
should be solved by the RG equation, or can be phenomenological
expressed in an effective potential \cite{Richard,Huangxg,JFXu15}.
In the infrared region at low energy, the dynamical gluon mass
represents the confinement feature of QCD \cite{Natale}.
Furthermore, in the presence of a strong magnetic field, the gluon
mass becomes large together with a decreasing of the interaction
constant, which leads to a damping of the chiral condensation. For
sufficiently strong magnetic fields $eB\gg\Lambda^2_\mathrm{QCD}$,
the coupling constant $\alpha_s$ is proposed to be related to the
magnetic field \cite{Miransky,Ferreira14}. Motivated by the work of
Miransky and Shovkovy \cite{Miransky}, the similar ansatz of the
magnetic-field-dependent coupling constant is introduced in the
SU(3) NJL models \cite{Ferreira14}. The simple ansatz of the running
coupling  is probably suitable for the SU(3) NJL model if we include
the $s$ quarks \cite{Ferreira14},
\begin{eqnarray}G'(eB)=\frac{G}{\ln(e+|eB|/\Lambda_\mathrm{QCD}^2)},\label{eq:G(eB)SU3}
\end{eqnarray} where the parameter $\Lambda_\mathrm{QCD}=300$ MeV.
We can find the running coupling constant versus the field $B$
approaches gradually to the constant value $G'(B\rightarrow 0)\sim
G$. In the computation of the SU(3) NJL model, we adopt the
parameters $\Lambda=602.3$ MeV, $m_u=m_d=5.5$ MeV, $m_s=140.7$~MeV,
$G=1.835/\Lambda^2$, and $K=12.36/\Lambda^5$  \cite{Rehber96}.

\section{Numerical Results and discussion }\label{sec:result}

\begin{figure}[ht]
\centering
\includegraphics[width=7cm,height=7cm]{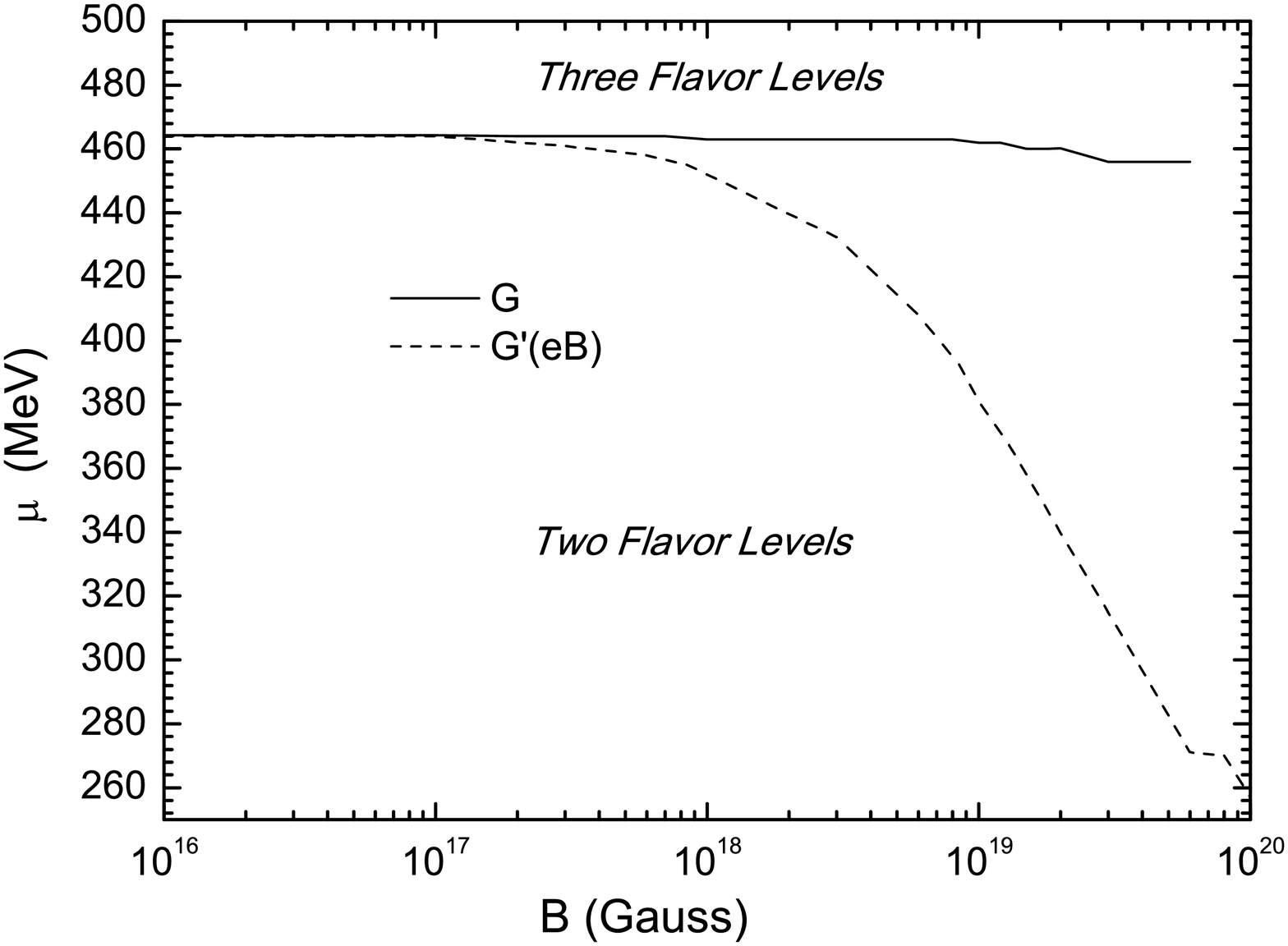}
\caption{The lines [solid and dashed lines for couplings $G$ and
$G'(eB)$ respectively] for the critical chemical potential $\mu$
separate the panel into two regions. In the region above the line,
there are strange quark distributions in Landau levels. In the
region below the line, only two-flavor light quarks fill in their
Landau levels, where the $s$ quark is excluded because of its
dressed mass larger than $\mu$.
        }
\label{fig1}
\end{figure}

In a strong magnetic field with a certain direction, quarks wrap
around the magnetic field and the orbital motion will be ruled by
the Landau energy level. Because dressed masses of quarks are
different, the occupations of the discretized Landau level are
flavor dependent. At a proper density, the dynamical masses of the
$u$ and $d$ quarks are smaller than their chemical potential and
have a real distribution in the Landau Levels. But the dynamical
mass of the $s$ quark is much heavier than that of the $u$/$d$
quark, and thus it cannot occur until the critical chemical
potential above its dynamical mass is reached. In Fig. \ref{fig1},
the critical chemical potential is shown by the solid line for the
conventional constant coupling $G$, and by the dashed line for the
running coupling $G'(eB)$ respectively. When the chemical potential
is above the line, the $s$ quarks have real distributions in the
Landau Levels. In the region below the line, there are only
two-flavor quarks in their Landau levels, and the $s$ quark is
excluded because of its dressed mass larger than $\mu$. It can be
found that at much higher field strengths, the regions of three
quark matter in the $\mu$-$B$ plane become wider. Furthermore, the
region is much wider with the running coupling $G'(eB)$ than the
constant coupling $G$. Therefore, it is concluded that the running
coupling interaction could broaden the region of three-flavor quark
matter. In other words, the strange quark can exist at lower density
with the running coupling than the constant coupling case.

\begin{table}[htbp]
\centering \caption{\label{tab:level} The quantum number of Landau
Levels occupied by quarks for the fixed chemical potential $\mu=350$
MeV at several magnetic fields. The number $``0"$ means the LLL.}
\begin{tabular}{cccccc}
\hline \hline Magnetic field (G) & \ $k_{u,\mathrm{max}}$   \ &\ $k_{d,\mathrm{max}}$ \  &\  $k_{s,\mathrm{max}}$ \  \\
\hline
$1.0\times 10^{17}$ &  115 & 230  & No   \\
$1.0\times 10^{18}$ &  15 & 30  & No   \\
$1.0\times 10^{19}$ &  1  & 3  & No   \\
$2.0\times 10^{19}$ &  0  & 1 & 0   \\
\hline \hline
\end{tabular}
\end{table}

It is well known that the chemical potential dominates the energy
spectrum of the particle without the magnetic field. Now we study
how large the magnetic field effect is on the distribution of Landau
levels at a fixed chemical potential. In Table \ref{tab:level}, we
adopt $\mu=350$ MeV, and give the maximum number of the Landau
levels of the $u$, $d$, and $s$ quarks for several magnetic fields.
It can be found that the much higher magnetic field can accommodate
quarks in lower levels. Furthermore, the onset of the $s$ quark can
be seen in its LLL until the field reaches $2\times 10^{19}$ G.
While at lower magnetic field, the quantum number of filled Landau
levels is larger and the quantization effects are washed out.

\begin{figure}[ht]
\centering
\includegraphics[width=7cm,height=7cm]{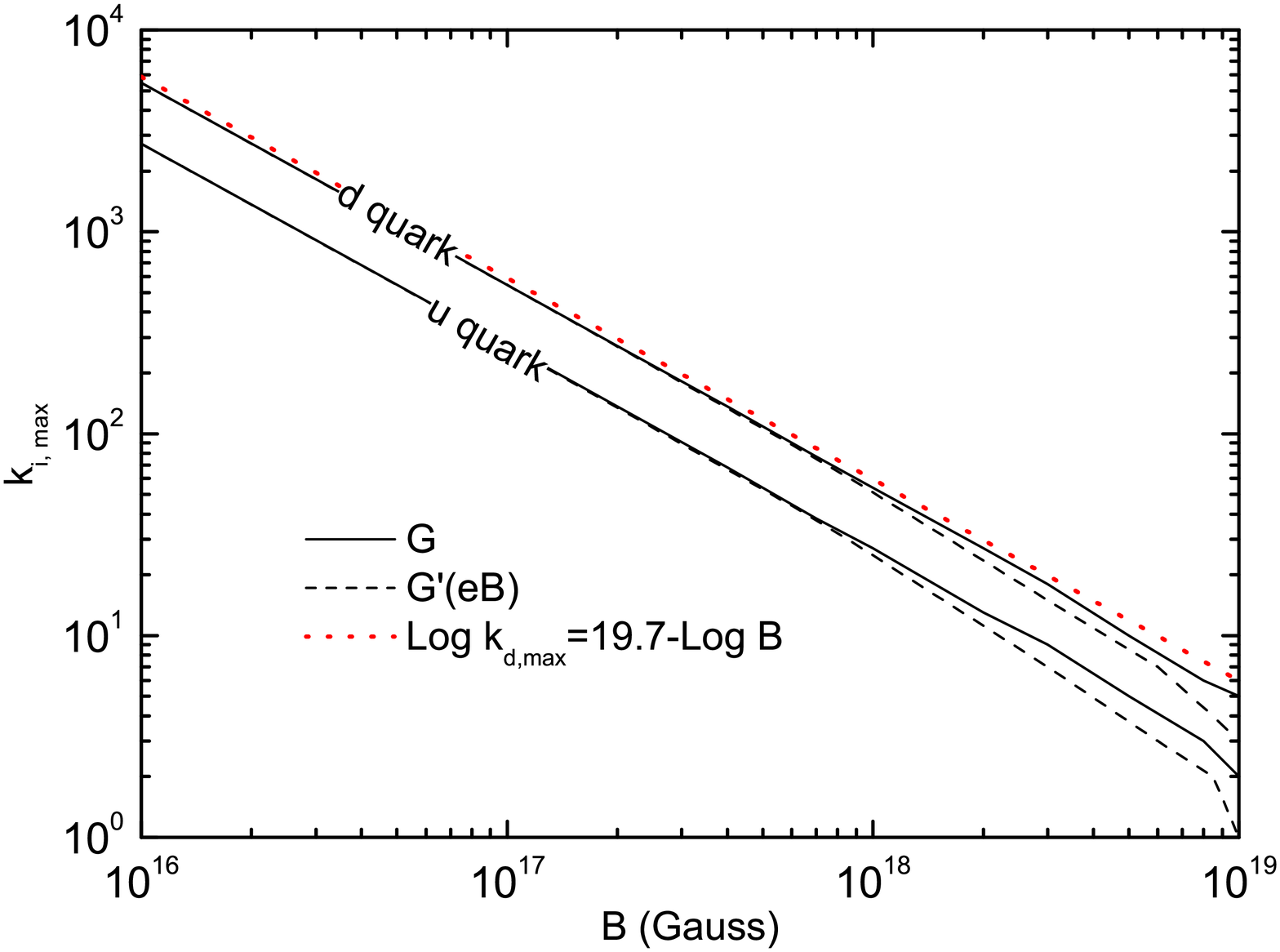}
\caption{The maximum Landau Levels for $u$ and $d$ quarks change
with the strong magnetic field. The solid line and dashed line are,
respectively, for the constant coupling $G$ and the running coupling
$G'(eB)$.
        }
\label{fig2}
\end{figure}
\begin{figure}[hbt]
\centering
\includegraphics[width=7cm,height=7cm]{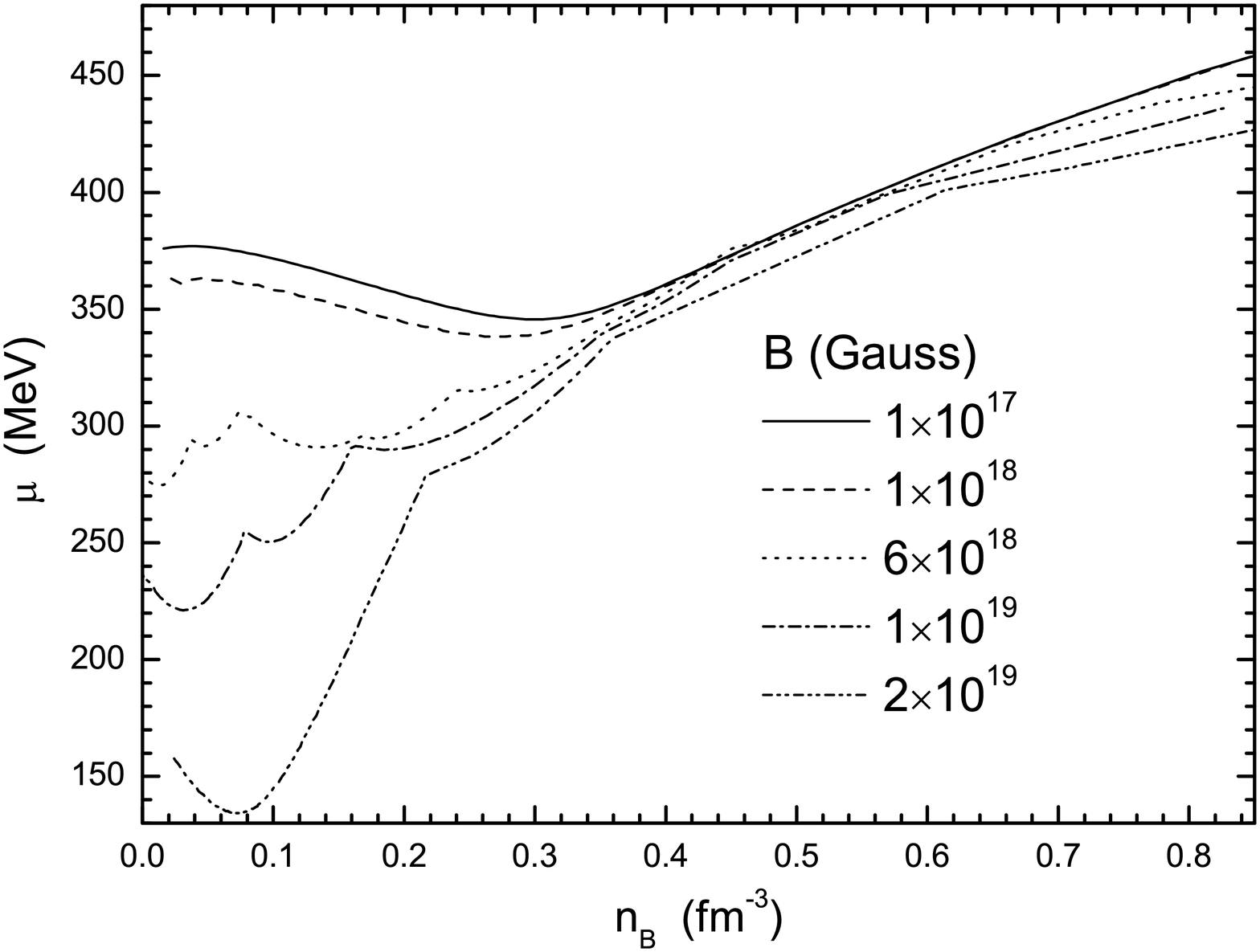}
\caption{ The chemical potential versus the baryon number density at
several values of the magnetic field.
        }
\label{fig3}
\end{figure}
Before the onset of the $s$ quark, the $u$ and $d$ quarks dominate
the quark matter. Because of the identity $q_u=-2 q_d$, the level
number of the $d$ quarks is exactly 2 times the level number of the
$u$ quark in order to meet the global charge neutrality. In Fig.
\ref{fig2}, the Landau Levels of the $u$ and $d$ quarks with the
same chemical potential $\mu$ are shown in the range of the magnetic
field ($10^{16}- 10^{19}$ G). We use the logarithm to label both the
vertical axis and the horizonal axis. Then the Landau level number
as functions of the magnetic field vary linearly, which is very near
the red dotted line ($\log k=19.7 -\log B$) for $d$ quarks.
Consequently, we can easily find that the $u$/$d$ quark Landau level
number $k_{i,\mathrm{max}}$ and the magnetic field $B$ nearly
satisfy a simple inverse proportional relation,
\begin{equation} k_{i,\mathrm{max}} \approx B_i^0/ B
\end{equation}
at the coupling $G$, where the scale is $B_d^0=5\times 10^{19}$ G
for $d$ quarks, which is 2 times $B_u^0$ of $u$ quarks. The
experiential formula indicates the constraint on the strong magnetic
field. In strong magnetic fields, charged fermions acquire infrared
phase space proportional to $|eB|$. As the magnetic field strength
increases, quarks are suppressed to the lower levels. At the same
time the degeneracy factor of each energy level is enlarged to $eB$.
When the magnetic field increases up to the order of $B_0$ or so,
almost all three flavors are concentrated on the LLL. The LLL would
make the QCD matter more interesting, where the quarks are
independent on the magnetic field with zero transverse energy \cite
{Kojo14}. After taking into account the running coupling $G'(eB)$
(marked by dashed lines) in Fig. \ref{fig2}, the line deviates from
the straight line at the field strength larger than $10^{18}$ G.
Therefore the running coupling can move the location of the LLL to a
field strength slightly lower than $B_0$. It can be expected that at
the same magnetic field, the SQM can be more easily realized at the
running coupling than the conventional coupling.

The strong magnetic field drastically affects the structural
properties and the thermodynamics. As far as we know, the chemical
potential increases together with the number density. But for the
SQM under a strong magnetic field, the variation relation between
the chemical potential and the number density is not always
monotonous. In Fig. \ref{fig3}, the quark chemical potential changes
with the baryon number density at several values of the magnetic
fields. The curves from top to down denote the increasing of the
magnetic field. At a low density less than $0.35$ fm$^{-3}$, the
effect of the magnetic field strength is very important, where the
degeneracy contribution from the magnetic field is much larger than
the fermion momentum. So at the same density, the chemical
potentials are very different for different magnetic fields.
Furthermore, a small number of quarks under the influence of the
strong fields can easily produce the oscillation behavior of the
chemical potential. While in the high density region, the fermion
momentum increases and the oscillation behavior cannot be easily
found anymore.

\begin{figure}[hbt]
\centering
\includegraphics[width=7cm,height=7cm]{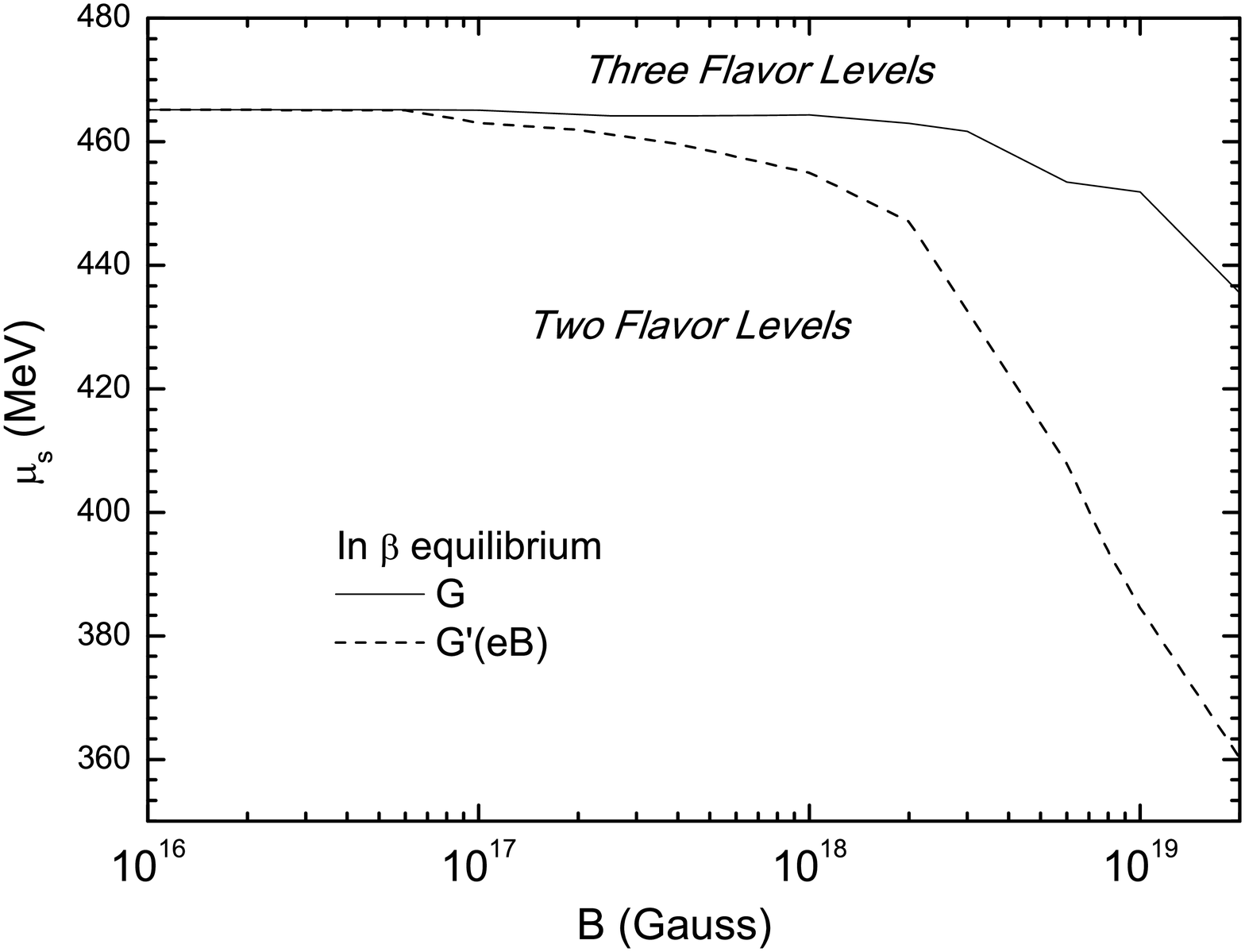}
\caption{ The critical chemical potential $\mu_s$ as the Fig.
\ref{fig1} is shown under the $\beta$ equilibrium condition.
        }
\label{fig40}
\end{figure}

In realistic situations for neutron stars, the chemical potentials
for different flavors will be different and related by the physical
constraints in a neutron star. So we can do the calculation by
assuming the three-flavor quark matter is in $\beta$ equilibrium.
Now there are three dynamical masses and two independent chemical
potentials, which can be determined by the three gap equations
(\ref{eq:gapSU3}), the baryon number conservation, and the neutral
charge condition,
\begin{eqnarray} 2n_u-n_d-n_s-3 n_e=0.
\end{eqnarray}

\begin{figure}[hbt]
\centering
\includegraphics[width=7cm,height=7cm]{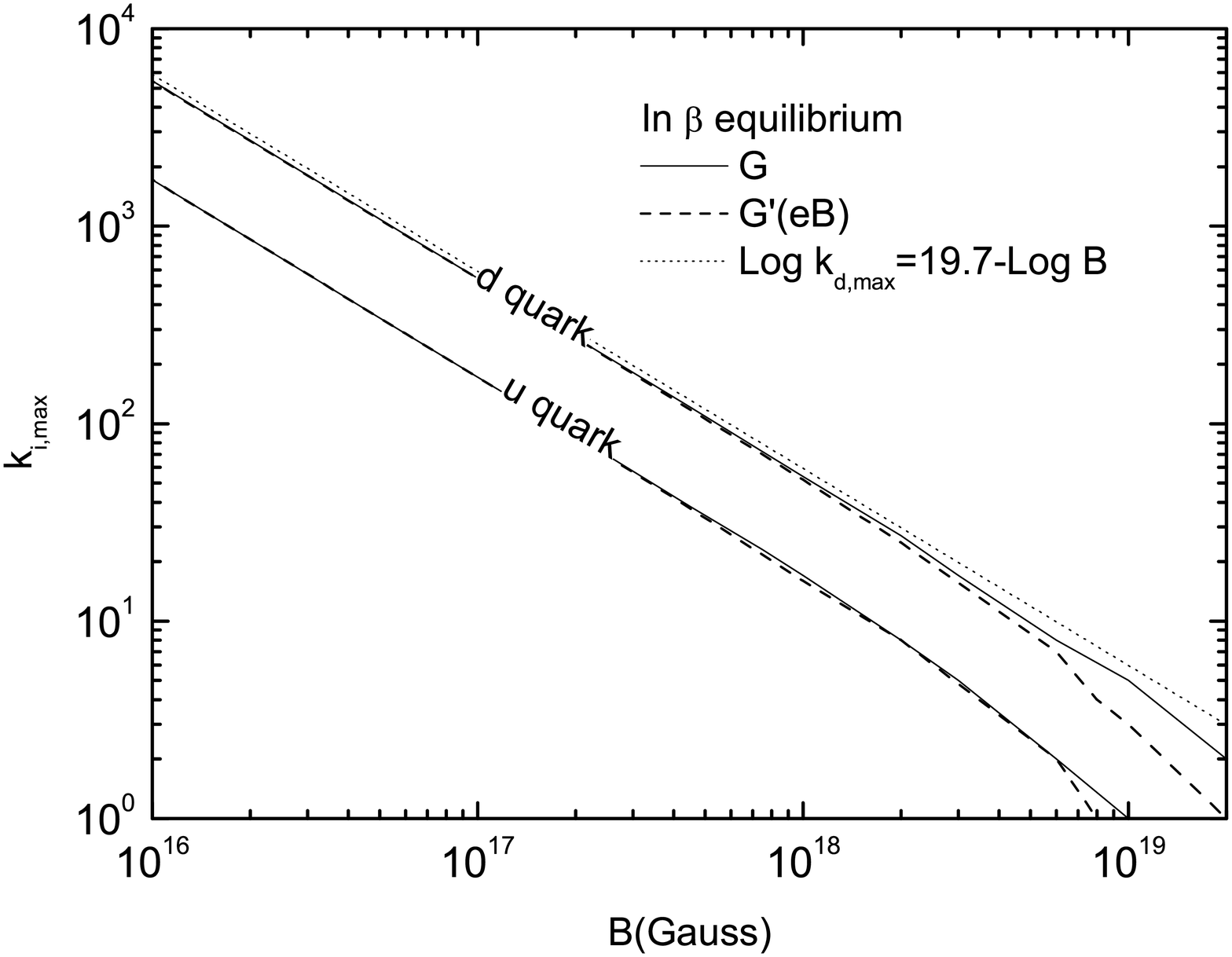}
\caption{ The maximum Landau Levels for $u$ and $d$ quarks versus
the magnetic field as in Fig. \ref{fig2} are shown under the $\beta$
equilibrium.
        }
\label{fig4}
\end{figure}

Under the $\beta$ equilibrium condition $\mu_d=\mu_s=\mu_u+\mu_e$ in
Fig. \ref{fig40}, we can get the similar result as in Fig.
\ref{fig1}, namely, the critical potential $\mu_s$ for the onset of
$s$ quark is about $465$ MeV with the coupling $G$. But at much
higher magnetic field, the value of $\mu_s$ has an apparent drop due
to the contribution of electrons.

\begin{figure}[hbt]
\centering
\includegraphics[width=7cm,height=7cm]{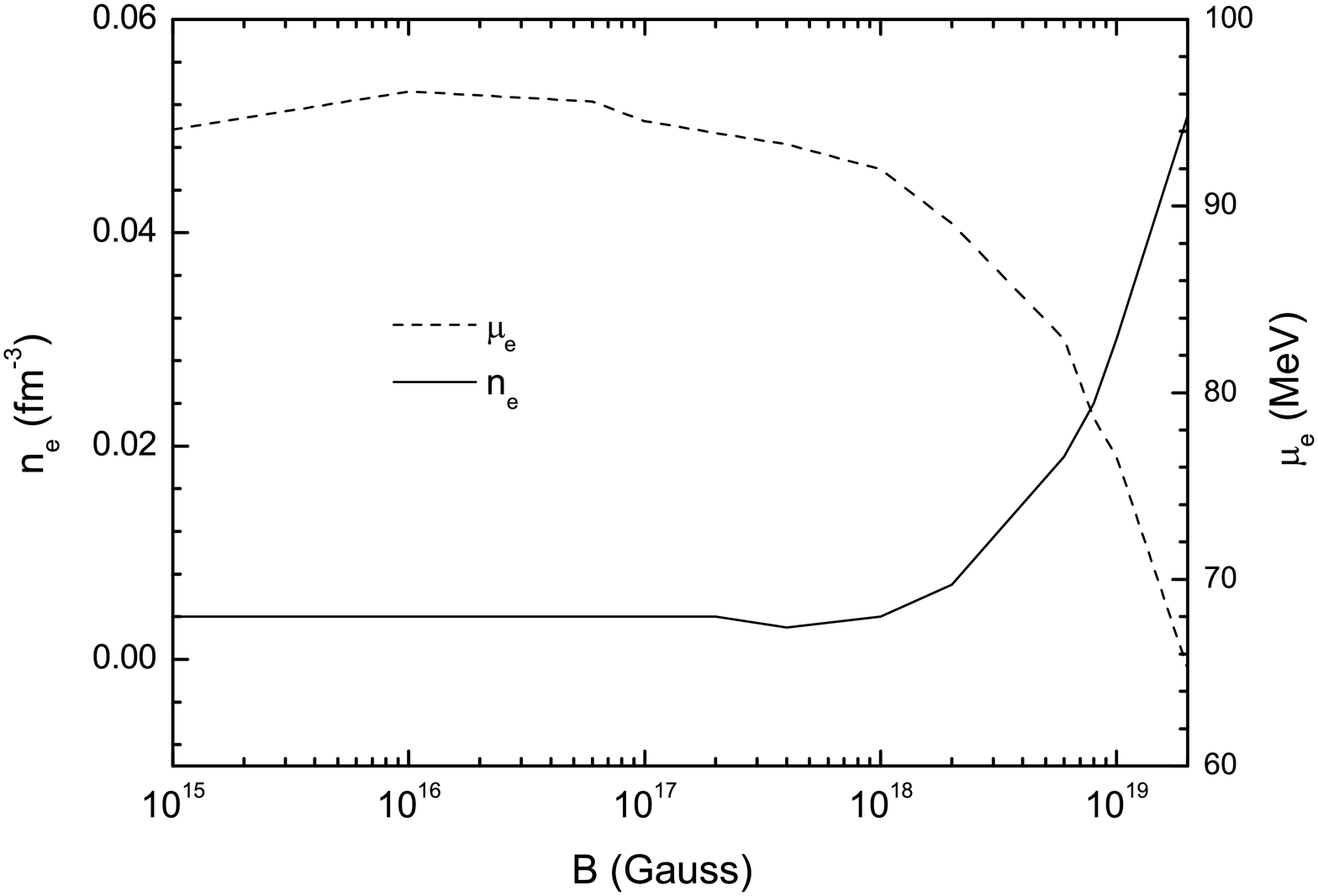}
\caption{ The density (on the left axis) and the chemical potential
(on the right axis) of electrons versus the magnetic field.
        }
\label{fig6}
\end{figure}

Before the onset of the $s$ quarks, the Landau level number is
approximately inverse proportional to the magnetic field. In Fig.
\ref{fig4}, we can also find the similar inverse proportional
relation between the maximum Landau quantum number and the magnetic
field. Under the $\beta$ equilibrium, it should be changed as
\begin{eqnarray} k^{\beta}_{i,\mathrm{max}} \approx B^{\beta}_i / B.
\end{eqnarray}
We can see that the contribution of electrons can hardly change the
relation of $d$ quarks. But the value of $B^{\beta}_u$ is decreased
to $1.7\times 10^{19}$ G because  its chemical potential is reduced
to $\mu_s-\mu_e$. In fact, the charge neutral condition can still be
reached due to the reduction of the $d$ quark density in addition to
the contribution of electrons. In Fig. \ref{fig6}, the corresponding
contribution of the electron density is shown on the left axis. The
chemical potential $\mu_e$ is shown on the right axis. We can see
that the $\mu_e$ will decrease as the magnetic field increases. On
the contrary, due to the degeneracy factor $eB$, the density will
keep increasing monotonously at the higher magnetic field.

\section{summary}
In this paper we have studied the energy level of the SQM in a
strong magnetic field within the SU(3) NJL model. The critical
chemical potential was shown for the onset of $s$ quarks. We found
that the running coupling scheme can broaden the existing region of
the three-flavor quarks in the $\mu$-$B$ plane. As the density
increases, the quarks in the high level participate in the system.
Just before the onset of $s$ quarks, we found that the maximum
quantum number of the $u/d$ Landau level is directly dependent on
the field strength through a simple inverse proportional relation
$k_{i,\mathrm{max}}\approx B_i^0/B$. The value of $B_0$ is certainly
flavor dependent. In particular, the value $B_d^0$ of $d$ quarks is
approximately 2 times $B_u^0$ of $u$ quarks. It should be pointed
that the chemical potential does not monotonously vary as the
density increases. At high densities, the chemical potential keeps
increasing with the density. But at low density, the chemical
potential could be a decreasing function of the baryon number
density, otherwise, the oscillation behavior becomes clear. The
corresponding work can be done under the $\beta$ equilibrium. The
running coupling can still broaden the window of three-flavor
matter. Because the common chemical potential relation is changed to
$\mu_d=\mu_s=\mu_u+\mu_e$, the inverse proportional relation should
be written as $k^{\beta}_{i,\mathrm{max}} \approx B^{\beta}_i / B$,
where $B^{\beta}_u$ is no longer half of $B^{\beta}_d$. These
results will be helpful in realistic situations for neutron stars.
The magnitude of magnetic field in compact stars decreases from the
core to the surface of stars. The study of the relation of magnetic
field and Landau level is helpful to know the components of stars
for a given radius.

The quarks in the LLL make the QCD more interesting and illustrate
the nonperturbative effects, which is further enhanced by the strong
magnetic field. We give the condition of the magnetic field larger
than $B_i^0$, under which all three-flavor quarks are lying in the
LLLs. We expect that all of these considerations would be helpful to
the theoretical investigation and future experiments searching for
the SQM under an extremely strong magnet field.

\begin{acknowledgments}
The authors would like to thank the National Natural Science
Foundation of China (No. 11475110, No. 11135011, and No. 11575190)
for support.

\end{acknowledgments}

\end{document}